\documentclass[reprint,amsmath,amssymb,aps,prc,nofootinbib]{revtex4-1}
\usepackage{bm}
\usepackage{graphicx}
\usepackage{dcolumn}
\usepackage{color}
\newcommand{\be}{\begin{equation}}
\newcommand{\ee}{\end{equation}}
\newcommand{\bea}{\begin{eqnarray}}
\newcommand{\eea}{\end{eqnarray}}

\newcommand{\mbsu}[1]{\mbox{\scriptsize #1}}

\newcommand{\vphu}{\vphantom{*}}

\newcommand{\ve}{\varepsilon}

\begin{document}

\title{Excitation spectra of exotic nuclei in a self-consistent
phonon-coupling model}

\author{N. Lyutorovich}
\author{V. Tselyaev}
\email{tselyaev@mail.ru}
\affiliation{St. Petersburg State University, St. Petersburg 199034, Russia}
\author{J. Speth}
\affiliation{Institut f\"ur Kernphysik, Forschungszentrum J\"ulich, D-52425 J\"ulich, Germany}
\author{P.-G. Reinhard}
\affiliation{Institut f\"ur Theoretische Physik II, Universit\"at Erlangen-N\"urnberg,
D-91058 Erlangen, Germany}

\date{\today}

\begin{abstract}
We explore giant resonance spectra and low-lying dipole strength in
the Ni and Sn chains from proton rich to very neutron rich isotopes,
relevant in astrophysical reaction chains. For the theoretical
description we employ the random-phase approximation (RPA) plus
many-body effects through a phonon coupling model with optimized
selection of phonons. The nuclear force is based on the
Skyrme-Hartree-Fock energy functional carried consistently through all
steps of modelling. The main effect of phonon coupling is a broadening
of the spectral distributions (collisional width).  This broadening is
particularly dramatic for low-lying dipole strength in very neutron
rich nuclei delivering there a qualitative change of the spectra.
\end{abstract}

\maketitle

\section{Introduction}
\label{sec:introduction}
For several decades, an extension of the random-phase approximation
(RPA) which includes the effects of the quasiparticle-phonon coupling
\cite{Soloviev_1992,Bertsch_1983,Kamerdzhiev_2004} has been
successfully applied in nuclear structure calculations.  The coupling
to phonons gives rise
to a broadening of the RPA states which is called collisional
  broadening \cite{Bertsch_1983} and which adds to the escape width
  from coupling to single-particle continuum.  An other very
important effect is the fragmentation of the bound collective
states. Due to this effect the low-lying collective $3^-$, $5^-$
  and $2^+$ states are shifted to lower energies without increasing
the corresponding BE-values \cite{Tselyaev_2017}. This solved a
problem in conventional RPA calculations which usually highly
overestimate the BE-values if one tries to reproduce the excitation
energies.  There exists many different versions of
quasiparticle-phonon coupling models.  They include the
quasiparticle-phonon model
\cite{Soloviev_1992,Vdovin_1983,Voronov_1983}, the core-coupling RPA
\cite{Krewald_1977}, the particle-vibration coupling model
\cite{Bertsch_1983,Colo_1994,Col01a,Roca-Maza_2017}, the two-phonon
extended RPA \cite{Barbieri_2003,Dickhoff_2004}, and the different
versions of the time-blocking approximation (TBA)
\cite{Tselyaev_1989,Kamerdzhiev_1993,Kamerdzhiev_2004,Tselyaev_2007,Litvinova_2007RTBA,Litvinova_2008,
  Lyutorovich_2015,Lyutorovich_2016,Tselyaev_2016,Tselyaev_2017,Tselyaev_2018}
developed within the many-body Green-functions formalism.  Here we use
the self-consistent version of the TBA developed in a series of our
recent papers
\cite{Lyutorovich_2015,Lyutorovich_2016,Tselyaev_2016,Tselyaev_2017,Tselyaev_2018}
where we used different approximations to solve the TBA equations
within this approach.

Although successful, phonon-coupling models leave some ambiguity
  concerning the choice of the number of RPA phonons which one
includes in the numerical approaches as the energy shift and the
fragmentation of the single-particle strength depends on this
number. As only a few of the RPA solutions represent collective states
and the majority of the solutions are more or less one-particle
one-hole ($ph$) states, a too large number of phonons would cause a
violation of the Pauli principle and it would introduce a severe
problem of double counting. Various measures to confine the
  selection of coupling phonons to ``collective'' ones had been
  considered in the past. We first selected the phonons according
to their reduced transition probability only considering states which
exhausted a certain fraction of the total transition strength.  Here
one assumes that the most collective states also couple most strongly
to the single-particle ($sp$) states which however, is not so
obvious. Recently we introduced a modified method of the TBA which
allows the selections of phonons in terms of dimensionless
  coupling strengths \cite{Tselyaev_2017} and for which we have
  demonstrated that it converges very fast and delivers clear cut
  criteria. Lately we developed a nonlinear version of our
  phonon-coupling model, an extended TBA, convergences automatically
  when enlarging the phonon space \cite{Tselyaev_2018}. With this
  extended TBA, we have counter-checked our previously developed
  cutoff criteria and found them confirmed.  As the new, extended TBA
  is much more computer time consuming we used the previous cutoff
  criterion in which the relevant phonons (positive frequency:
  $\omega^{\vphu}_{n} > 0$) are chosen according to
  \cite{Tselyaev_2017}
\begin{subequations}
\be
  |\,v^{\vphu}_n\,| > v_{\mbsu{min}}
  \;,\;
  v^{\vphu}_n =
  \langle\,V\,\rangle^{\vphu}_n / \omega^{\vphu}_{n}
  \;,
\label{vndef}
\ee
where $\langle\,V\,\rangle^{\vphu}_n $ represents the average residual
interaction in a given RPA state $|\,{z}^{n} \rangle$ with the energy
$\omega_n$.  This means that one considers only those phonons whose
dimensionless interaction strength $v_n$ is larger than the
cutoff value $v_{\mbsu{min}}$.  The mean value of the residual
interaction is connected with the energy shift of a given RPA state
\be
\langle\,V\,\rangle^{\vphu}_n = |\,\omega^{\vphu}_{n}| -
|\,\omega^{(0)}_{n}|\,.
\label{vmnf}
\ee
where
\be
|\,\omega^{(0)}_{n}| = \sum_{ph}\,(\ve^{\vphu}_{p}-\ve^{\vphu}_{h})
\bigl(\,|\,z^{n}_{ph}\,|^2 + |\,z^{n}_{hp}\,|^2 \bigr)\,.
\label{omznph}
\ee
\end{subequations}
is the average single-particle energy of state $n$.
  Collective states are distinguished by a strong shift in
energy. Therefore, this criterion picks preferably collective
  states as desired. The success depends on choosing a proper values
  for the cutoff parameter $v_{\mbsu{min}}=0.05$. In
Ref. \cite{Tselyaev_2017} it has been demonstrated that stable
  results are obtained for the parameter $v_{\mbsu{min}}$ in the range
  0.01--0.1 in which the numerical results remain nearly unchanged.
In the actual calculations we will chose $v_{\mbsu{min}}=0.05$.

In the present publication, we apply TBA to two chains of doubly
magic nuclei ranging deep into the regime of exotic nuclei
relevant for astro-physical reaction chains \cite{Arnould_2007}.
We calculated excitation energies and transition probabilities of bound
states as well as dipole strength distributions in the regime of
giant resonances for four Ni ($^{48,56,68,78}$Ni) isotopes and four Sn
($^{100,132,140,176}$Sn) isotopes. Both considered isotope chains are
unique in the nuclear chart because in all these cases there exists a
certain chance to obtain experimental numbers for some of the
low-lying collective states and giant resonances. For the more
  stable members of the chains, experimental values do already exist
  and will be referred to later on. We expect reliable results as our
theoretical model has been successfully tested on several doubly magic
nuclei, including $^{56}$Ni and $^{132}$Sn from the present
  chains.

Furthermore, we have a look at the low-lying electric dipole
strength in these nuclei. Although we do not find particularly
collective states in this region (in accordance with
\cite{Reinhard_2013}), we denote it here by the nickname pygmy dipole
resonance (PDR) as commonly used in the literature. The structural
changes in the PDR region along the isotopes are particularly
interesting because our chains span from proton rich isotopes to very
neutron rich ones.

We ought to mention that there exist numerous theoretical
publications where aspects of our investigation have been
addressed. We try to summarize these references restricting ourselves
to Ni and Sn isotopes. The question of a possible magic neutron number
$N=40$ has been discussed in Refs. \cite{Yao_2015,Saxena_2017}. This
is important in connection with the ``magic'' nucleus $^{68}$Ni which
we investigate in our paper. For this isotope, several experimental
data exist: (I) the PDR \cite{Wieland_2009,Wieland_2011} and (II) the
isoscalar giant monopole resonance (GMR) \cite{Vandebrouck_2014}.
The absense of the low-lying pigmy resonance in the $^{68}$Ni
monopole response function is discussed in \cite{Hamamoto_2014}.

In many of the papers the giant dipole resonance (GDR) and the low-lying
dipole strength have been addressed.  In the proton rich $^{48}$Ni
isotope, the PDR has been identified as a vibration of loosely bound
protons against the proton-neutron symmetric core. Here two different
nuclear structure models have been used \cite{Paar_2005}. Within a
self-consistent quasiparticle RPA the low-lying dipole strength has
been calculated \cite{Papakonstantinou_2014,Papakonstantinou_2015_err}
in a number of Sn isotopes using the Gogny force and in a similar
approach also Ni isotopes were investigated
\cite{Papakonstantinou_2015}. In Ref.~\cite{Co_2013}, the PDR was
studied in various medium and heavy nuclei with self-consistent RPA
models using also the Gogny interaction.  Here, a detailed comparison
of the PDR and the conventional GDR was given. The E1 strength for
fifteen Sn isotopes haven been calculated in self-consistent models
including pairing correlations \cite{Avdeenkov_2011}. The authors also
highlighted the astrophysical aspect of their investigation. The
energy of the $2^+_1$ state in $^{68,78}$Ni and other Ni isotopes were
calculated in Ref.~\cite{Hagen_2016} in the framework of the
coupled-cluster theory with chiral nucleon-nucleon and three-nucleon
interactions. The continuum was taken into account by employing the
Berggreen basis which treats bound-, resonant-, and non-resonant
scattering states on equal footing. The predicted range for the
$2^+_1$ state in $^{78}$Ni is considerably higher than for its
neighbors which considered the authors as an indication that this
nucleus might be doubly magic.  Quasiparticle-vibration coupling
effects on nuclear transitions of astrophysical interest were studied
in \cite{Robin_2017} where the relativistic version of the
quasiparticle time-blocking approximation was used.  PDR strength in
Ni isotopes was studied in Ref.~\cite{Sun_2018} using a deformed RPA.
This confirmed the predominantly isoscalar character of the PDR and
discussed the relation between PDR and neutron skin.  Low-lying
isoscalar and isovector dipole strength was investigated in
Ref. \cite{Kim_2016} for $^{48}$Ni and other $N=20$ isotones. Larger
amounts of E1 strength in the asymmetric $N = 20$ isotones were
predicted than in their $Z = 20$ mirror nuclei.  Low-lying states in
$^{100}$Sn and neighboring nuclei were calculated in
\cite{Morris_2018} within an {\it ab initio} approach using realistic
interactions.  The relation of various giant resonance properties with
the symmetry energy has been reviewed in the overview article
\cite{Colo_2014}.

\begin{table*}
\caption{\label{tab:ground-state} Ground state properties
of the nuclei under consideration.
All energies are in MeV, $T_{1/2}$ units are given in the table.}
\begin{ruledtabular}
\begin{tabular}{c|ccc|cc|cc}
      & \multicolumn{3}{c|}{Experiment}
                                  & \multicolumn{2}{c|}{HF(SV-bas)}
                                                 & \multicolumn{2}{c}{HF(SV-m64k6)}
\\
     &$T_{1/2}$& $S(n)$& $S(p)$   & $\epsilon_F(n)$
                                          & $\epsilon_F(p)$
                                                 & $\epsilon_F(n)$
                                                        & $\epsilon_F(p)$
\\
\hline
      &        &       &          &       &      &      &
\\
$^{48}$Ni
      & $2.1^{+1.4}_{-0.6}$ ms \cite{Pomorski_2014}
               &  --   &  --      & -22.40&+0.065&-23.35&-0.0105
\\
$^{56}$Ni
      & 6.075(10) d
               & 16.639& 7.165    &-15.49 &-6.35 &-16.08& -6.84
\\
$^{68}$Ni
      & 29(2) s& 7.792(5)& 15.431(7)
                                  & -8.18 &-14.19&-8.40 &-14.69
\\
$^{78}$Ni
      & $0.11^{+0.10}_{-0.06}$ s
               & 5.5     &  --    & -5.27 &-20.18&-5.59 &-20.93
\\
$^{100}$Sn
    &1.16(20) s& 17.410  & 3.200  &-16.39 &-2.49 &-17.02& -2.96
\\
$^{132}$Sn
     &39.7(8) s& 7.311 & 15.710   &-7.22  &-14.96&-7.82 &-15.47
\\
$^{140}$Sn
      &  --    &  --   & --       & -2.64 &-16.88&-2.49 &-17.03
\\
$^{176}$Sn
      &  --    &  --   &   --     & -1.07 &-26.16&-1.79 &-26.32
\\
\end{tabular}
\end{ruledtabular}
\end{table*}

The paper is organized as follows.  In Sec.~\ref{sec:formal}, we
briefly outline our approach.  In Sec.~\ref{ground}, we discuss the
ground-state properties of the Ni and Sn isotopes under consideration.
The calculated mean energies and widths of giant resonances in the
nuclei are discussed in Sec.~\ref{sec:GR-global}.  A more detailed
description of giant resonances in terms of their strength
distributions is given in Sec.~\ref{sec:GR-detail}.  In
Sec.~\ref{sec:PDR}, the fine structure of the PDR is analyzed.
Conclusions are given in the last section.  The appendix contains
numerical details of our calculations.

\section{Formal framework}
\label{sec:formal}

\subsection{The underlying mean-field model: Skyrme-Hartree-Fock}
\label{sec:SHF}

Basis of the description is the successful and widely used
Skyrme-Hartree-Fock (SHF) functional.  It is an energy-density
functional depending on a couple of local densities and currents
(density, gradient of density, kinetic-energy density, spin-orbit
density, current, spin density, kinetic spin-density). It is augmented
by a density-dependent pairing functional. However, the present survey
deals only with doubly-magic nuclei in which pairing is not active.  A
detailed description of the functional, its calibration, and its
properties is found in the reviews
\cite{Bender_2003,Stone_2007,Erler_2011}.  It is important to note
that SHF functionals cannot yet be derived with satisfying precision
from {\it ab initio} calculations. All available parametrizations are
determined by fits to experimental data, in some cases with additional
constraints on nuclear matter properties.  Different groups fit with
different data sets and preferences. Thus there exists a great variety
of parametrizations, most of them performing comparably well in basic
nuclear ground-state properties.  Larger variances are found in the
prediction of excitations, particularly in the isovector channel.  For
this survey, concentrating on giant resonance spectra, we select two
parametrizations where the description of the GDR was taken into
account in the construction. The first one is SV-bas from
\cite{Kluepfel_2009} which manages to provide, within RPA, a pertinent
description of four response properties in $^{208}$Pb, namely dipole
polarizability, GDR, GMR, and isoscalar giant quadrupole resonance
(GQR). The second parametrization under consideration is SV-m64k6 from
\cite{Lyutorovich_2012} which was fitted to the same data set as
SV-bas but with the aim to reproduce, within TBA, the GDR in
$^{208}$Pb and $^{16}$O. These both parametrizations have considerably
different nuclear matter properties, effective mass
$m^*/m(\mbox{SV-bas})=0.9$ versus $m^*/m(\mbox{SV-m64k6})=0.64$, TRK
sum rule enhancement ($\equiv$ isovector effective mass)
$\kappa(\mbox{SV-bas})=0.4$ versus $\kappa(\mbox{SV-m64k6})=0.6$,
and symmetry energy $J(\mbox{SV-bas})=30$ MeV versus
$J(\mbox{SV-m64k6})=27$ MeV. They thus provide a rough indicator of
the variance of extrapolations resonance spectra in exotic nuclei.

\subsection{Phonon-coupling in time-blocking approximation (TBA)}
\label{sec:TBA}

We use the fully self-consistent phonon-coupling model built on top of
RPA using the SHF functional in the particular form as we developed
and refined it in a series of previous publications
\cite{Lyutorovich_2015,Lyutorovich_2016,Tselyaev_2016,Tselyaev_2017}.
Both, RPA and TBA, use the same numerical representation. The $sp$
energies and $sp$ wave-functions are obtained from stationary
Skyrme-Hartree-Fock (SHF) calculations and the residual $ph$
interaction for RPA and TBA is computed fully self-consistently using
the same SHF functional~\cite{Lyutorovich_2016} (ignoring, however,
spin modes which play no role here).  RPA and TBA are evaluated with
proper treatment of the nucleon continuum from \cite{Tselyaev_2016}
while the phonons entering the effective interaction in TBA were
computed in a discrete basis.  The effective interaction in TBA is
renormalized by subtraction of the zero-frequency component to stay
consistent with the mean-field ground state and to render the TBA
solutions stable \cite{Toe88a,Tselyaev_2013}.  The selection of
phonons in TBA was optimized according to inverse energy
weighted strength \cite{Tselyaev_2017}.  Actually, the phonon basis
was chosen with the cutoff parameter $v_{\mbsu{min}}=0.05$ and a
maximal phonon energy $E^{\mbsu{phon}}_{\mbsu{max}}$=40 MeV.

The numerical solution exploits spherical symmetry and uses a
coordinate space representation on a radial grid with grid spacing
0.05 fm and box size 18 fm, for a detailed discussion of box size see
appendix \ref{app:box}. Another crucial numerical parameters is the
space of $sp$ states.  The maximal angular momentum of the $sp$ basis
was $l^{\mbsu{sp}}_{\mbsu{max}}$ =17 which has proven to be sufficient
in all calculations so far.  The choice of maximum $sp$ energy
$\ve^\mathrm{sp}_\mathrm{max}$ depends on the application.  For giant
resonance spectra with their rather coarse energy smoothing of order
of 100 keV, we take $\ve^\mathrm{sp}_\mathrm{max}=100$ MeV.  Low lying
dipole strength in the PDR region was evaluated with finer
energy resolution which, in turn, requires a higher energy
cutoff. Here we take, after careful tests of convergence,
$\ve^\mathrm{sp}_\mathrm{max}=200$ MeV. For details of the choice of
$sp$ space see appendix \ref{app:sp-basis}.

Although the present TBA treatment includes two crucial broadening
mechanisms in detail (escape width due to continuum description,
collisional width by phonon coupling), the spectral distributions of
the giant resonance are still plagued by some residual fluctuations
because the phonon input to TBA stays yet at the level of discrete RPA
and employs a limited basis (missing, e.g., higher configurations as
$4p4h$ etc).  We thus employ an additional smoothing by Lorentzian
effectively generated by augmenting the energy in the propagator
denominator by a small imaginary part $i\Delta$.  For the gross
structures of giant resonance, we use folding with 400 keV width and
for the more detailed PDR (near and below continuum threshold) only 10
keV.

\section{Results}

\subsection{Ground-state properties}
\label{ground}
%
Table~\ref{tab:ground-state} shows some ground state properties of the
exotic nuclei under consideration: half-lives $T_{1/2}$, experimental
proton and neutron separation energies $S(p)$ and $S(n)$, compared
with calculated Fermi energies $\epsilon_F(p)$ and
$\epsilon_F(n)$.
The dominant process limiting lifetime is
$\beta$-decay, except for $^{48}$Ni which has a considerable
branching to proton emission \cite{Pomorski_2014}.

The separation energies are in the average
fairly well reproduced by both SHF parametrizations and they are all
positive, with possible exception of $^{48}$Ni which may be a proton
emitter. Thus all nuclei (except $^{48}$Ni) used in the following are
still stable against nucleon emission. However, the nucleon emission
threshold can be extremely low which means that the low-lying fraction
of fragmented resonance spectra lies in the nucleon continuum. This
will become particularly important for the low-lying dipole strength,
the PDR branch.

\subsection{Giant resonances - global properties}
\label{sec:GR-global}

\begin{figure}\centering\footnotesize
\includegraphics[width=\linewidth]{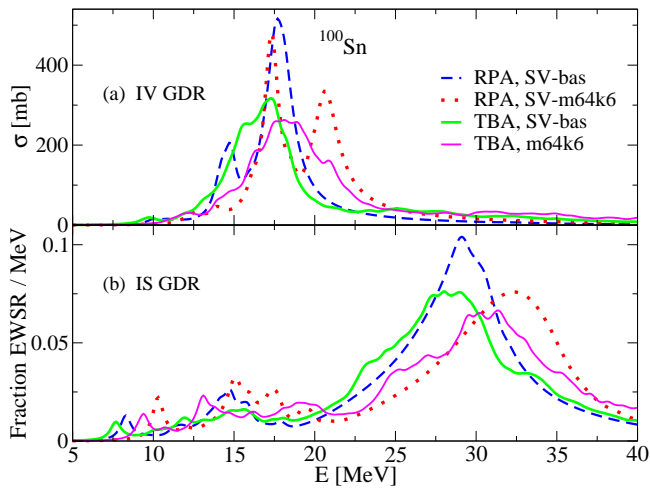}
    \caption{\label{fig:100Sn_IV+IS_GDR_rpa+tba_400keV} Comparing
  photo-absorption spectrum
and the spectrum of the isoscalar E1 excitations
  from TBA and RPA for the case of
  $^{100}$Sn computed with the two parametrizations SV-bas and
  SV-m64k6 and smoothing of $\Delta=400$ keV.  }
\end{figure}
In previous publications, we have studied extensively giant resonance
spectra in stable, doubly magic nuclei, for a systematic survey see
\cite{Tselyaev_2016}. Two general features can be extracted. First,
TBA produces some small downshift of the average resonance energy as
compared to RPA. Second, and more important, the main effect of TBA is
a significant smoothing of the often rather spiky RPA spectra
delivering realistic spectral
profiles. Figure~\ref{fig:100Sn_IV+IS_GDR_rpa+tba_400keV} serves to
demonstrate that for the example of one exotic nucleus $^{100}$Sn. We
consider here the GDR because this is a mode with strong spectral
fragmentation where the beneficial effect of smoothing by TBA can be
well seen. The RPA results show for both parametrizations strong
fragmentation peaks which are so well separated that even the a
posterior smoothing with $\Delta=400$ keV cannot wipe out the
(unphysical) structures. TBA resolves that problem at once delivering
TBA spectra with the typical GDR profile of one broad resonance
peak. The key to success is that TBA induces effectively an
energy-dependent smoothing because the phase space of phonon
configurations increases with excitation energy.

The basic giant resonance, IS GMR and IS GQR as well as IV GDR, can be
characterized by two numbers, the average peak position and the width
of the resonance distribution.  Resonance peak positions are evaluated
as the energy centroid $E_0=m_1/m_0$ of the first and zeroth energy
moments of the corresponding strength distributions $S(E)$ in a given
energy window around the maximum. For the IV GDR (here we considered
the photo absorption cross section) as well as for the IS GMR and GQR
(here we considered multipole strengths),
the energy windows are $E_0 \pm 2\delta $ where
$E_0$ is the peak position and $\delta$ is
the spectral dispersion.  To avoid too small energy windows, we used
the constraint $\delta > \delta_{\text{min}}$ with
$\delta_{\text{min}}=3$ MeV for the GMR, \ 2.5 MeV for the GQR, and
\ 4 MeV for the IV GDR.  The width $\Gamma$ and dispersion for these
resonances were defined as
\begin{equation}
  \Gamma = 2\delta \sqrt{2\ln 2}\,,
  \qquad
  \delta^2 = \frac{\int (E-E_0)^2\, S(E)\, dE}{m_0}
  \quad.
\end{equation}
The IS GDR and the PDR (low-lying dipole) strength have too complex
spectra and cannot be quantified that easily. Here we cannot avoid to
look at the full spectral distributions as done in sections
\ref{sec:GR-detail} and \ref{sec:PDR}.

\begin{figure}\centering\footnotesize
\includegraphics[width=\linewidth]{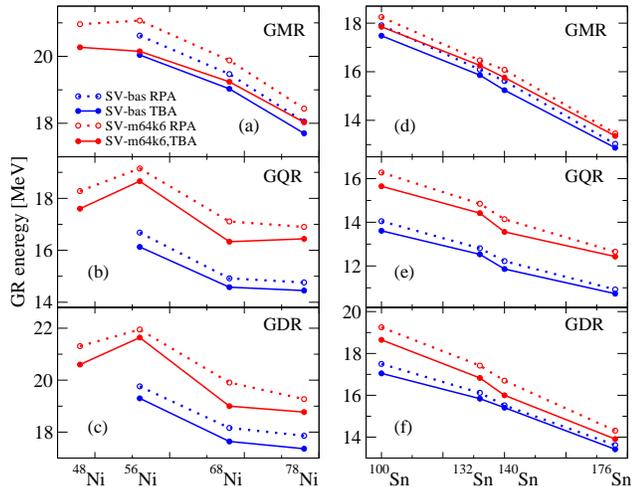}
\caption{\label{fig:GR_positions} Mean giant resonance energies (in
  units of MeV) calculated for the parameter sets SV-bas (circles
  connected with blue lines) and SV-m64k6 (diamonds connected with red
  lines). The RPA results are shown with dashed line and the TBA
  results with solid lines.  Results for Ni isotopes are shown in the
  left panels (a), (b), and (c) and those for Sn isotopes in the right
  panels (d), (e), and (f).  Upper panels (a) and (d) shows GMR,
  middle panels (b) and (e) GQR, and lower panels (c) and (f) GDR.}
\end{figure}
Fig.~\ref{fig:GR_positions} summarizes the average peak positions for
the three basic giant resonances and the two isotopic cases considered
in this paper. The effects are very similar to those observed
previously in stable nuclei \cite{Tselyaev_2016}.  With exception of
the outlier $^{48}$Ni, we see a smooth trend approximately
$\propto A^{-1/3}$ as typical for most giant resonances. The trend is
particularly well visible in the long Sn chain. We also see the
systematic difference between SV-bas and SV-m64k6 for the GQR energies
which is exclusively due to the different effective mass
\cite{Bra85aR,Kluepfel_2009}. This feature is not affected by TBA.
There is an interesting slight difference between SV-bas and SV-m64k6
in the trend of the GDR. Recall that SV-m64k6 was designed to overcome
\cite{Lyutorovich_2012} the wrong $A$-dependence of GDR predictions
seen in most Skyrme parametrizations \cite{Erl10a}. This modified
$A$-dependence is well visible already in the small section shown
here.

The effect of phonon coupling is generally a small down-shift by
phonon coupling of order 0--1 MeV, more for the lighter Ni nuclei and
less for the heavier Sn series. No systematic difference between the
three modes and no significant dependence on the parametrization can
be seen. These typical patterns were also seen for stable nuclei
\cite{Tselyaev_2016}. Thus far there is nothing particular is
happening for this selection of exotic nuclei, even at the extremes
of proton or neutron binding.

\begin{figure}\centering\footnotesize
\includegraphics[width=\linewidth]{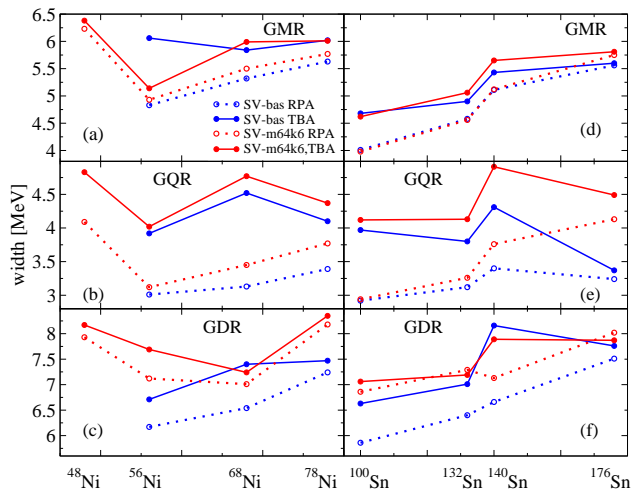}
\caption{\label{fig:GR_peaksandwidths} As Fig.~\ref{fig:GR_positions},
  but for the average widths of giant resonances.}
\end{figure}
Fig.~\ref{fig:GR_peaksandwidths} complements the previous figure by
showing the widths of the three resonances. Here the effects of phonon
coupling are, of course, relatively stronger and lead to increase of
typically 1 MeV. Not visible in this one number is the fact that this
extra width is predominantly used to smooth the still peaky pattern
of RPA spectra as we have seen in figure
\ref{fig:100Sn_IV+IS_GDR_rpa+tba_400keV}.  It is worthwhile to
note that an (energy dependent) folding with average width of 1 MeV is
often used in drawing RPA results to simulate smoothing by complex
configurations, see e.g. \cite{Nes02a}.

In case of widths, we can see some general trends (though with some
exceptions): 1. Widths tend to increase along a chain with increasing
neutron numbers; exception are extremely proton rich nuclei (example
$^{48}$Ni) where high density of proton states also drives larger
widths.  2. SV-m64k6 with its broader spectral stretch (low effective
mass) produces in RPA consistently larger widths than SV-bas; however,
this trend can be overruled by phonon coupling in TBA.  3. Widths are
generally increased by phonon coupling; exception are extremely
neutron rich isotopes where we see practically no increase of width by
phonon coupling.  The reason is here that already the RPA spectra are
very broadly distributed due to the high density of neutron states in
these very exotic nuclei. We will look at that in detail in the next
section.

\subsection{Giant resonances - detailed distributions}
\label{sec:GR-detail}

\begin{figure}\centering\footnotesize
\parbox[t]{8.5cm}{\centering
\includegraphics[width=8.5cm]{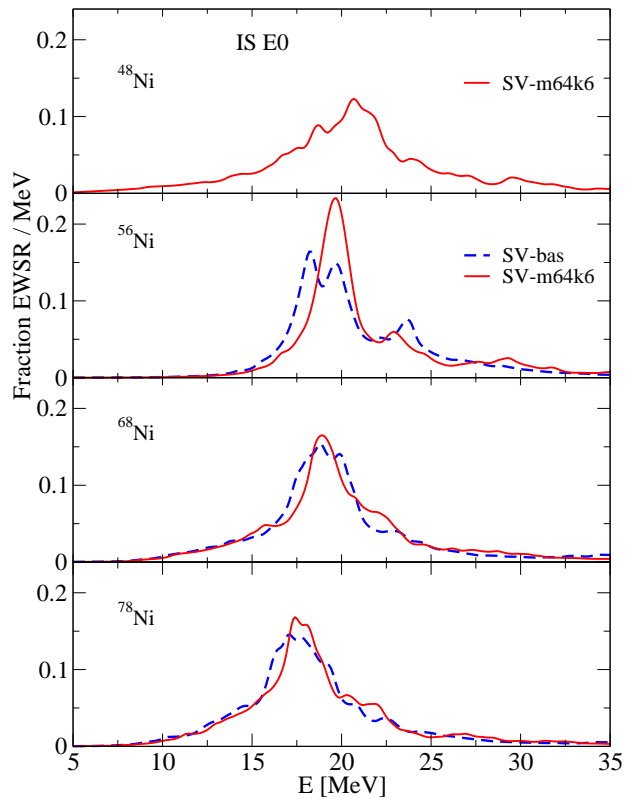}
\caption{\label{fig:Ni_ISE0_400keV} Isoscalar E0 strength (in units of
  fractions of the EWSR) in the Ni isotopes calculated in the TBA with
  $\Delta=400$ keV. The results for the parameter sets SV-bas and
  SV-m64k6 are shown by blue dashed and red full lines, respectively.
}}
\end{figure}
Figure \ref{fig:Ni_ISE0_400keV} shows the strength distribution of the
isoscalar GMR for the doubly magic Ni isotopes, $^{48,56,68,78}$Ni,
given in units of fractions of energy-weighted sum rule (EWSR) per
energy interval. The general structure is similar to what was seen in
stable nuclei, mostly one broad peak.  There is little difference
between SV-bas and SV-m64k6 which is plausible because the position of
the GMR is uniquely related to the incompressibility
\cite{Boh79aR,Bla80aR,Kluepfel_2009} and both parametrizations here have about
the same incompressibility.  The exotic nuclei provide a large span of
isospin which allow to see more clearly trends with neutron number, or
system size respectively. In this case, we see that the resonance
width is rather large at the proton-rich side and shrinks by almost
factor two for the neutron-rich isotopes. The peak position roughly
follows the $\propto A^{-1/3}$ law known from fluid dynamical models
of the GMR. We ought
to remind at this place that the global trend of GMR peak positions
seems not yet well under control with available SHF functionals
\cite{Kva16a}. The problem may be to some extend at the experimental
side because GMR are extremely hard to identify
unambiguously. Measurements of GMR in the exotic nuclei studied here
are even more demanding and will probably not show up soon.

\begin{figure}\centering\footnotesize
\parbox[t]{8.5cm}{\centering
\includegraphics[width=8.5cm]{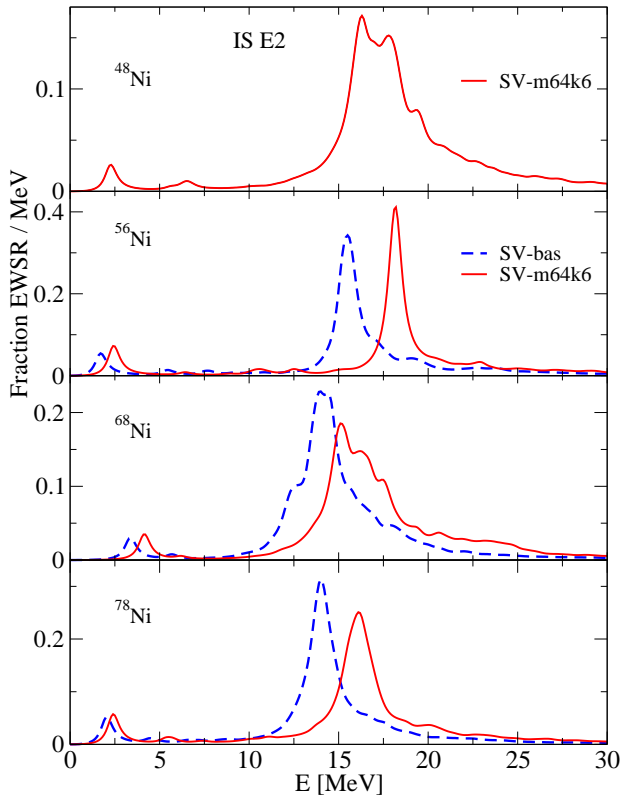}
\caption{\label{fig:Ni_ISE2_400keV}
Same as in Fig.~\ref{fig:Ni_ISE0_400keV} but for the isoscalar E2 strength.}
}
\end{figure}
Figure \ref{fig:Ni_ISE2_400keV} shows the results for the isoscalar
GQR.  This delivers similar pattern as seen for the GMR: one broad
peak for each isotope and widths shrinking with increasing neutron
number.  The widths are generally somewhat smaller than for GMR, a
relation which is already known from GQR and GMR in stable nuclei. For
the GQR, we see a significant difference in the predictions from
SV-bas and SV-m64k6. This, again, can be explained by the collective
properties of the GQR which is known to depend sensitively and almost
exclusively on the isoscalar effective mass
\cite{Bra85aR,Kluepfel_2009}. Recall that the effective mass of
SV-m64k6 ($m^*/m=0.64$) is much lower than that of SV-bas ($m^*/m=0.9$).

\begin{figure}\centering\footnotesize
\parbox[t]{8.5cm}{\centering
\includegraphics[width=8.5cm]{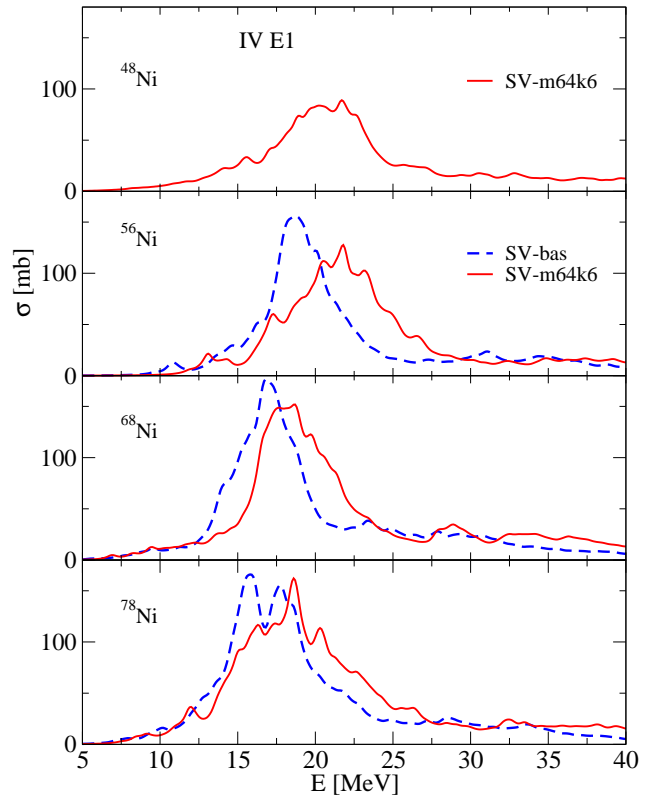}
\caption{\label{fig:Ni_photo_400keV}
Same as in Fig.~\ref{fig:Ni_ISE0_400keV} but for the photo-absorption
cross section, proportional to the energy weighted isovector dipole
strength.}
}
\end{figure}
Figure~\ref{fig:Ni_photo_400keV} shows photo-absorption cross sections
for the isovector GDR in Ni isotopes.  This mode is visibly more
fragmented than GMR and GQR, particularly with its long tails at high
and low energies. The latter correspond to the PDR region discussed
later on in section \ref{sec:PDR}. The bulk of the resonance is again
represented by one broad peak. The distribution is broader than for
GMR and GQR corresponding to a larger fragmentation in the GDR
channel. The detailed fragmentation structure, still dominating the
RPA strengths (not shown here), is wiped out by the higher
configurations modeled in TBA. The comparison between SV-bas and
SV-m64k6 shows an interesting effect: the average peak positions
display a different isotopic trend. There is no unique agent for that
because many parameters had been played with in SV-m64k4 to enforce
simultaneous description of the GDR in $^{16}$O and $^{208}$Pb
\cite{Lyutorovich_2012} (which is a general problem in SHF functionals
\cite{Erl10a}). But it is most likely that these different isotopic
trends are the main source arranging reasonable results for the GDR in
$^{16}$O (for the early presentation see \cite{Lyutorovich_2012} and
for results at the latest stage of model development
\cite{Tselyaev_2016}). The mechanisms could be worked out more deeply
with studying trends of GDR on long isotopic chains. This calls for
new GDR data on long chains which would be of invaluable help for
fixing this particular isovector aspect of SHF functionals. Chances
for such data are better than for GMR and GQR because the GDR is a
dominating channel in many reactions and because precision
measurements in the GDR region have become fashionable recently, see
e.g. \cite{Tamii_2011,Has15a}.

\begin{figure}\centering\footnotesize
\hfil
\parbox[t]{8.5cm}{\centering
\includegraphics[width=8.5cm]{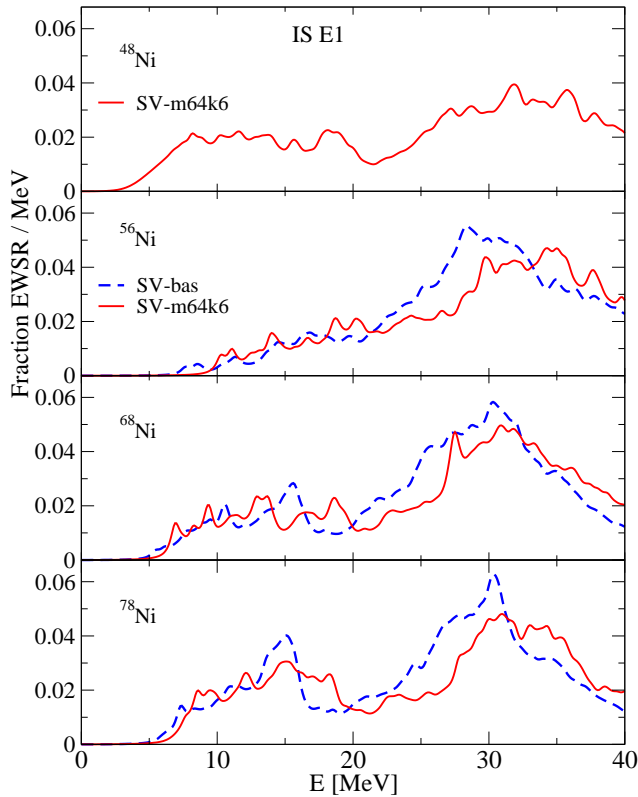}
\caption{\label{fig:Ni_ISE1_400keV}
Same as in Fig.~\ref{fig:Ni_ISE0_400keV} but for the isoscalar E1 strength.}
}
\end{figure}

Figs. \ref{fig:Ni_ISE0_400keV} and \ref{fig:Ni_ISE2_400keV} display
fractions EWSR for GMR and GQR, and Fig. \ref{fig:Ni_ISE1_400keV} the
photo-absorption strengths for the IS GDR. The EDF parameter sets
SV-bas and SV-m64k6 were used in all of these calculations, with the
exception of $^{48}$Ni. Since HF calculation with the set SV-bas does
not give a bound ground state for $^{48}$Ni (see
Table~\ref{tab:ground-state}), only the set SV-m64k6 was used in the
TBA calculations for this nucleus.

\begin{figure*}[p]\centering\footnotesize
\parbox[t]{8.5cm}{\centering
\includegraphics[width=8.5cm]{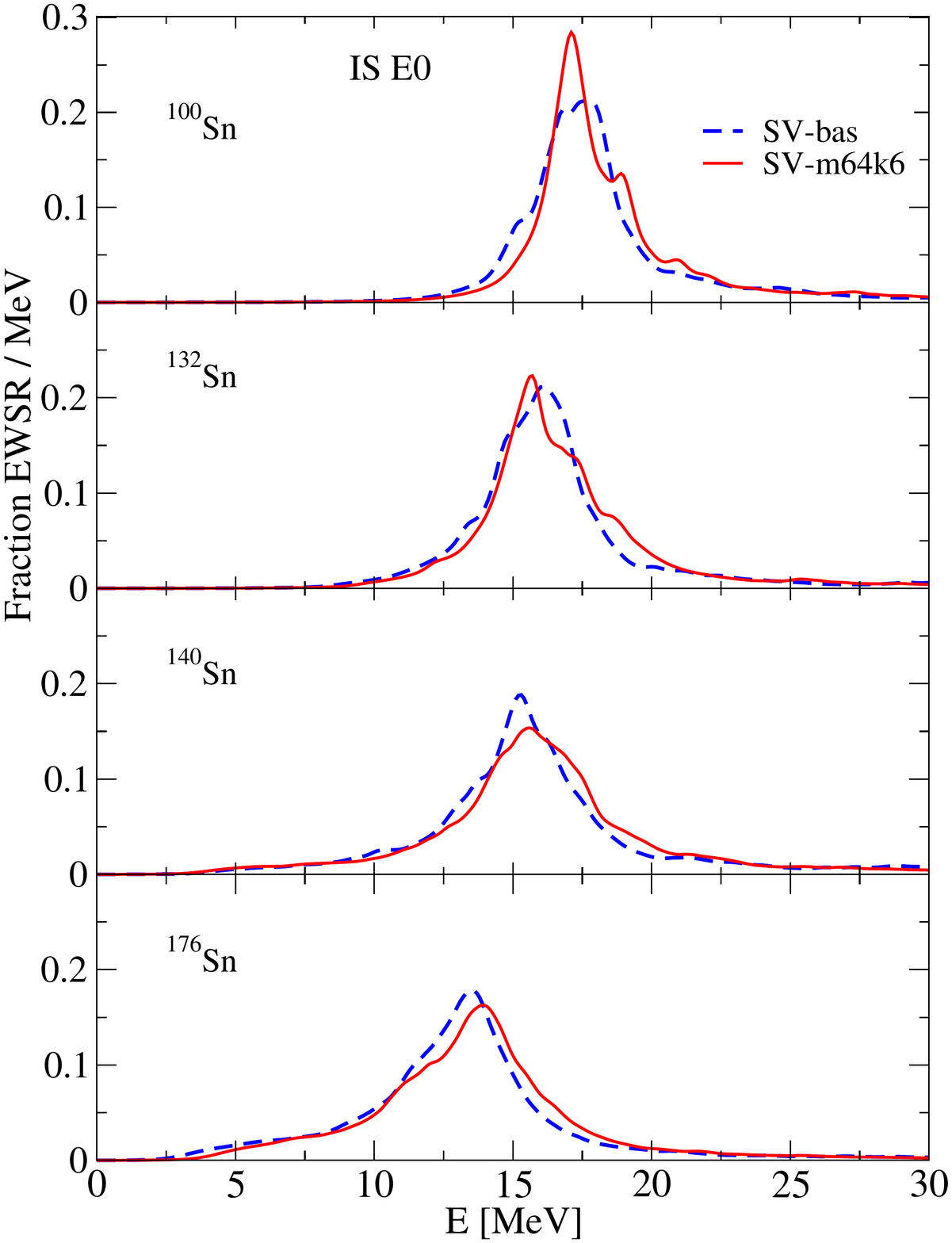}
\caption{\label{fig:Sn_ISE0_400keV}
Same as in Fig.~\ref{fig:Ni_ISE0_400keV} but for the IS E0 strength in
the Sn isotopes.}
}\hfil
\parbox[t]{8.5cm}{\centering
\includegraphics[width=8.5cm]{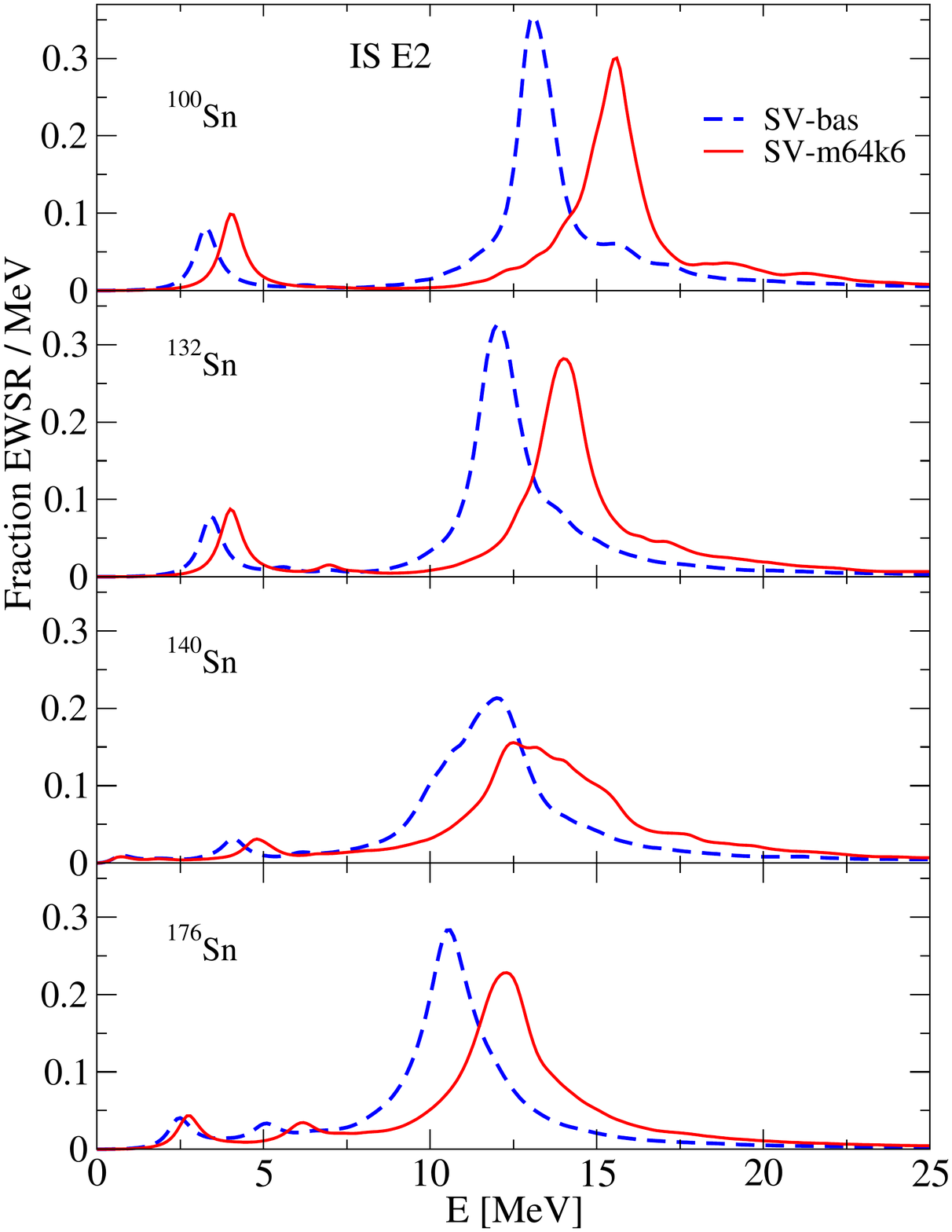}
\caption{\label{fig:Sn_ISE2_400keV}
Same as in Fig.~\ref{fig:Ni_ISE0_400keV} but for the IS E2 strength in
the Sn isotopes.}
}
\parbox[t]{8.5cm}{\centering
\includegraphics[width=8.5cm]{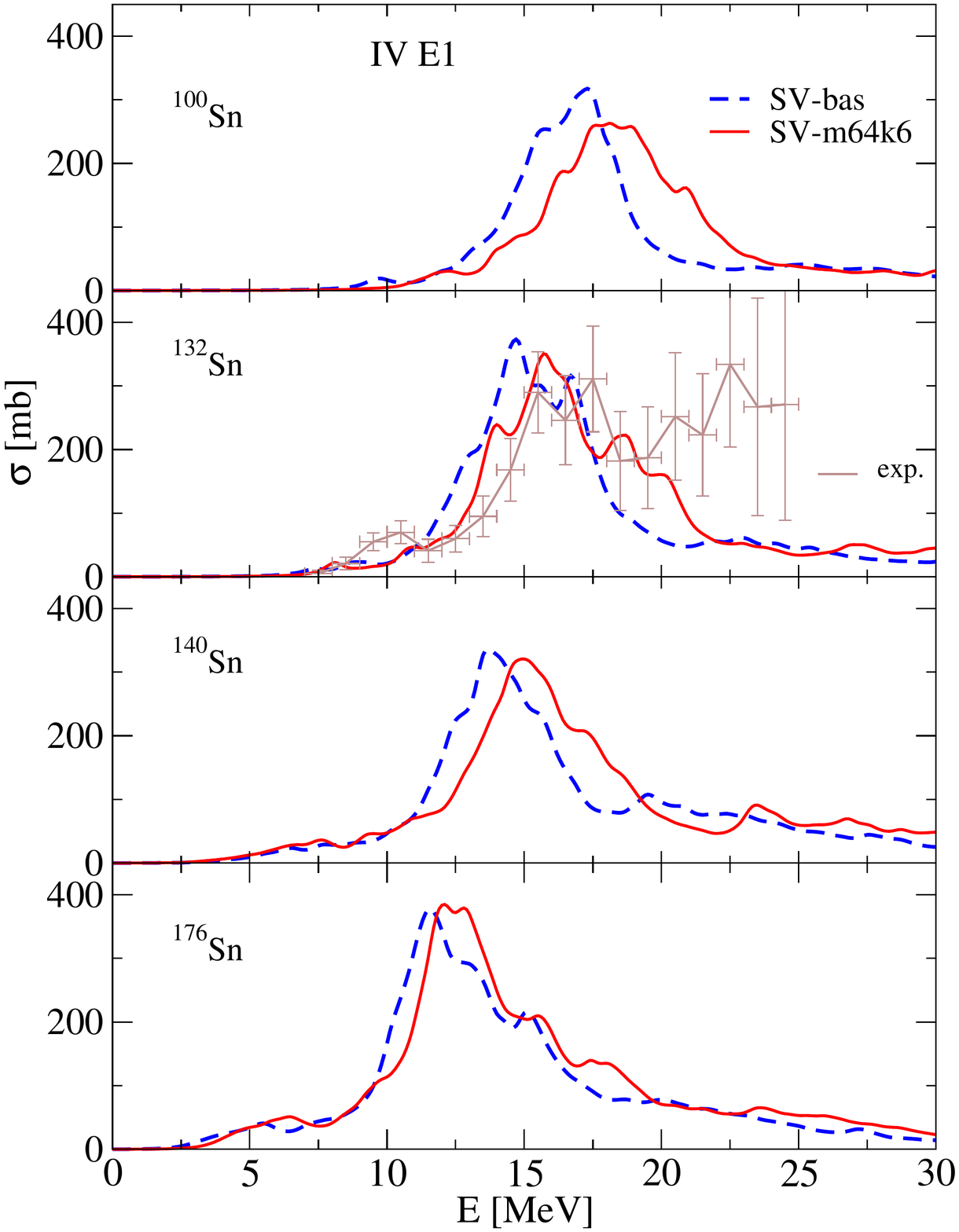}
\caption{\label{fig:Sn_photo_400keV}
Same as in Fig.~\ref{fig:Ni_ISE0_400keV} but for the photo-absorption
cross section in the Sn isotopes. The experimental data for $^{132}$Sn
\cite{Adrich_2005} are shown by the brown line with error bars.}
}\hfil
\parbox[t]{8.5cm}{\centering
\includegraphics[width=8.5cm]{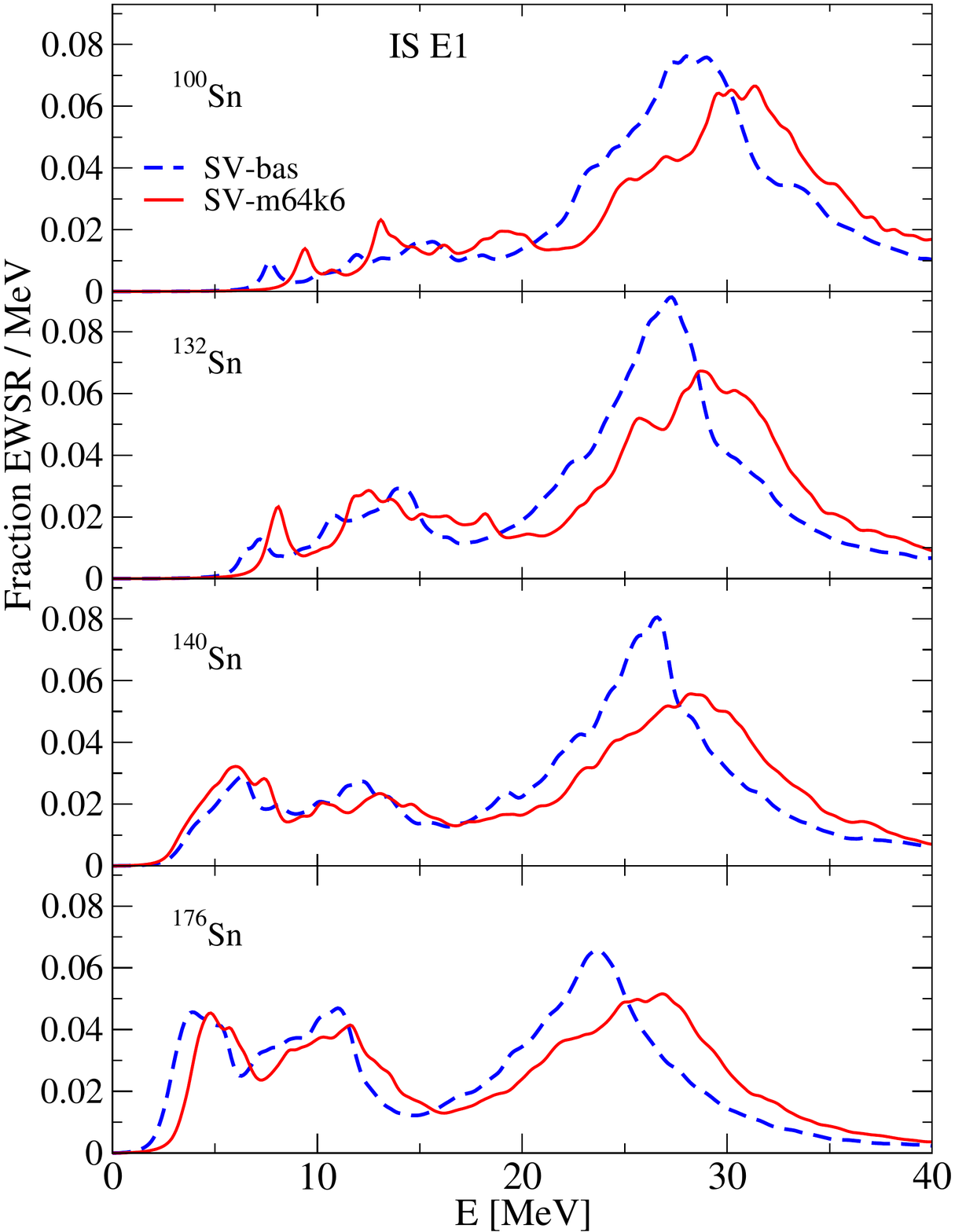}
\caption{\label{fig:Sn_ISE1_400keV}
Same as in Fig.~\ref{fig:Ni_ISE0_400keV} but for the IS E1 strength in
the Sn isotopes.}
}
\end{figure*}
As the last example in the series for Ni isotopes, we look at the
isoscalar dipole strength.  This channel explores two interesting
nuclear resonances, the compressional dipole mode at higher energies
and the toroidal mode at lower energies \cite{Har01aB}. The latter one
overlaps energetically with the PDR region and there is considerable
cross talk between these two modes \cite{Rep13a,Rep17a} which makes a
comparative discussion of isoscalar and isovector dipole strength
extremely useful for an understanding of these low-energy dipole
modes.  Figure \ref{fig:Ni_ISE1_400keV} shows the isoscalar dipole
strength distributions for Ni isotopes computed with TBA.  These
strength distributions are the most widely spread of all channels
discussed in this paper. The spectral fragmentation is so strong here
that RPA spectra looked already as smooth as the TBA spectra shown
here. There is no qualitative change of pattern nor a change of widths
with the isotopes which is understandable because this channel covers
two resonances which roughly maintain their position and fill the
spectrum in between by fragmentation. Particularly interesting is the
low-energy side, the toroidal branch, due to its cross-talk with PDR
and its high sensitivity to low-lying particle-hole excitations.  The
latter feature is seen here from the fact that the low-energy branch of
isoscalar dipole strength shows strong dependence on neutron number in
the series here. The same happens, in fact, also for the low-lying
isovector strength which we will see when zooming into that region in
section \ref{sec:PDR}.

The same series of resonance channels is shown for the chain of
doubly-magic Sn isotopes $^{100,132,140,176}$Sn in
Figs.~\ref{fig:Sn_ISE0_400keV}, \ref{fig:Sn_ISE2_400keV},
\ref{fig:Sn_photo_400keV},
and \ref{fig:Sn_ISE1_400keV}.  Pattern and trends are
the same as in case of Ni isotopes. There is quantitative change in
pattern to the extend that the widths for GMR, GQR, and GDR are all
smaller than the corresponding widths for Ni. This is a known effect:
the heavier the nucleus the more concentrated the giant resonances.

Experimental data are indicated for the IV GDR in $^{132}$Sn.  They
were obtained by Coulomb dissociation of secondary Sn beams with
energies around 500 MeV/nucleon \cite{Adrich_2005}.  There is much
strength at the side of higher energies which is probably unrealistic
(note the large error bars).  No theoretical model shows such pattern,
not only ours here but also other calculations with RPA and beyond
\cite{Tsoneva_2008,Robin_2017,Yueksel_2012}. Besides this dramatic,
but probably negligible, mismatch, we see that the main peak is well
reproduced by SV-m64k6 and approximately by SV-bas. What is not
fitting so well for both parametrizations is the pronounced low-energy
peak in the data. Qualitatively, the predictions are pertinent: there
is a low-lying peak, however slightly lower than data and, more
important, covering less strength. The example shows that data on
low-energy dipole strength (PDR region) in neutron rich nuclei provide
fruitful challenges for existing nuclear models.

\subsection{Low-lying dipole strength}
\label{sec:PDR}

\begin{figure}\centering\footnotesize
\includegraphics[width=\linewidth]{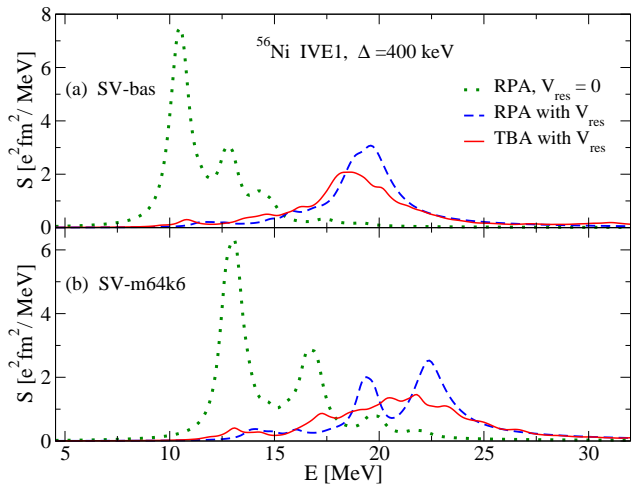}
\caption{\label{fig:V_residual} $^{56}$Ni IVE1 uncorrelated strength function
(green dotted line) in comparison with the RPA (blue dashed) and TBA
(red solid) results. Panels (a) and (b) show the results obtained with
SV-bas and SV-m64k6 forces.}
\end{figure}

As mentioned above, the isovector GDR strengths have remarkably long
tails.  Of particular interest is the low-energy tail, often called
PDR. In fact, it was shown that this region, although often producing
a satellite of dipole strength does not represent a single resonance,
but rather a collection of $sp$ strengths of mixed nature (dipole,
toroidal, compressional dipole) \cite{Rep13a,Rep17a,Rei13c} and as
such it may be even more interesting. The $sp$ dominated, mixed
structure leads to much higher sensitivity to details of the
model. Thus we have a look at both, RPA and TBA, in comparison. Before
we go into details we show in Fig. \ref{fig:V_residual} the
uncorrelated dipole strength function together with the RPA and TBA
results.  The sensitivity on the nuclear structure models can be seen
from the uncorrelated E1 strength distributions calculated with the
parameter set SV-bas and SV-m64k6, respectively. In both cases we see
the well known effect of the residual interaction which shift the
major part of the low-lying strength nearly 10 MeV higher creating the
GDR.  The phonons included in the TBA give rise to a small shift
downwards and increase the width.  The width of $\Delta$ = 400 keV is
too large to see differences between RPA and TBA in the PDR region.
Therefore we have to analyze the spectra with higher resolution. Here
it helps that these low energy states experience only small broadening
effects (escape width and collision width are naturally smaller
there). Thus the calculations for the PDR region were performed with
very small energy step 2.5 keV and small folding width $\Delta$ = 10
keV. Proper handling of nucleon continuum, as we do, is essential in
these calculations near threshold.  As the presentation becomes now
more detailed, we have to select a subset of nuclei. For Ni isotopes
(see figure \ref{fig:Ni_photo_400keV}), we concentrate on $^{56}$Ni as
proton rich isotope and $^{68}$Ni as neutron rich isotope where in
both cases dipole spectra display a marked low-energy tail. For Sn
isotopes (see figure \ref{fig:Sn_photo_400keV}), we consider
$^{100}$Sn as proton-rich example and $^{140}$Sn as a very
neutron-rich exotic nucleus.
\begin{figure}\centering\footnotesize
\includegraphics[width=\linewidth]{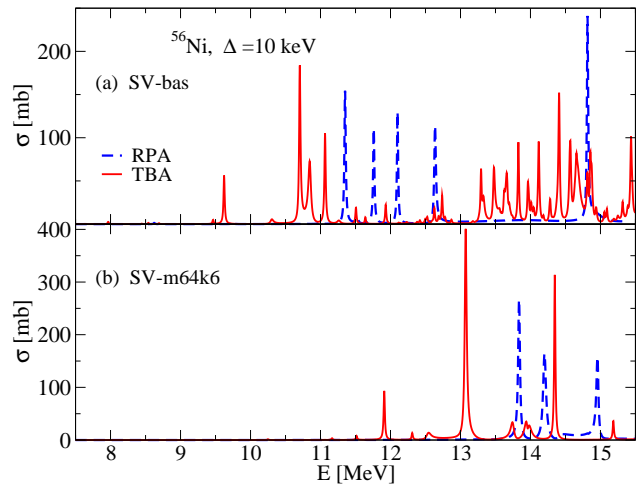}
\caption{\label{fig:56Ni_photo_delta10keV} Fine structure of
  low-energy isovector dipole strength in $^{56}$Ni: photo-absorption
  cross section $\sigma (E)$ calculated in RPA (blue, dashed) and TBA
  (red, solid) using $\Delta=10$ keV. Results for the two parameter
  sets SV-bas and SV-m64k6 are displayed in separate panels for better
  discrimination.}
\end{figure}

Figure \ref{fig:56Ni_photo_delta10keV} shows the low-energy wing of
isovector dipole strength in $^{56}$Ni.  The lower end of the PDR
spectra shows predominantly a down-shift and little
fragmentation. This is due to the fact that it has a rather large
HOMO-LUMO gap and therefore only very few low lying $ph$ states
which, in turn, provides too few low-energy phonons. The higher part
of the PDR branch (above 12 MeV) shows already some smoothing by
fragmentation. It is more pronounced for SV-bas because of its higher
spectral density.

\begin{figure}\centering\footnotesize
\parbox[t]{8.5cm}{\centering
\includegraphics[width=8.5cm]{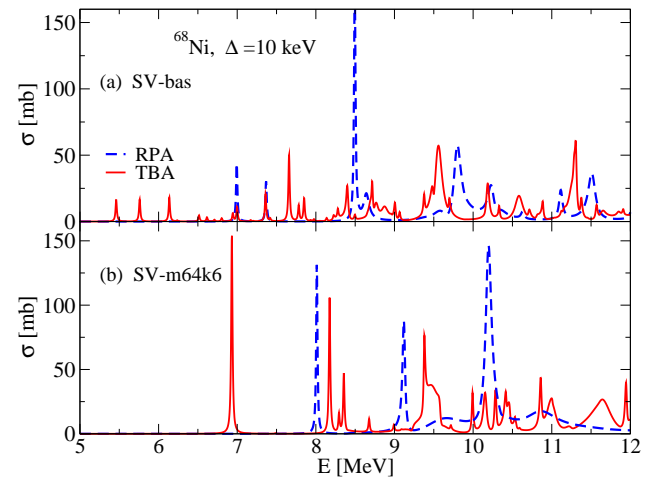}
\caption{\label{fig:68Ni_photo_delta10keV}
Same as in Fig.~\ref{fig:56Ni_photo_delta10keV} but for $^{68}$Ni.}  }
\end{figure}
Figure \ref{fig:68Ni_photo_delta10keV} shows the PDR part of the
isovector strength for the neutron rich $^{68}$Ni.  Due to the weak
neutron binding, this isotope has a long tail of low-lying neutron
$ph$ states and correspondingly a couple of low-lying phonons.  As a
consequence, we see in TBA substantial spectral fragmentation of the
isolated RPA peaks, again, more with SV-bas than with SV-m64k6.  The
higher resolution with which we scan the PDR spectra allows to track
in detail how TBA distributes the peaks of the RPA spectrum over
groups of smaller peaks.  There are states which basically survive
with some small down-shift in energy while others are practically
dissolved and spread over the neighborhood. It is not possible to
establish simple rules for that. We would also not recommend to take
the details literally, peak by peak. Mind that we underestimate the
effects of higher configurations which is probably not fully
compensated for by the only 10 keV folding width which we use here
merely to look at the spectra with higher resolution.  Robust, thus
more reliable, information are spectral densities and global
properties of the distribution. And here we see marked differences
between the two parametrizations. The distribution reaches deeper down
with SV-bas than with SV-m64k6. This effect as such is established
already at the level of RPA. In fact, it can be traced back to the
density of mere $ph$ states. TBA modifies only details as further
spectral fragmentation and possibly small energy shifts. Details
which, however, should be thought of when aiming at detailed analysis.

\begin{figure}\centering\footnotesize
\parbox[t]{8.5cm}{\centering
\includegraphics[width=8.5cm]{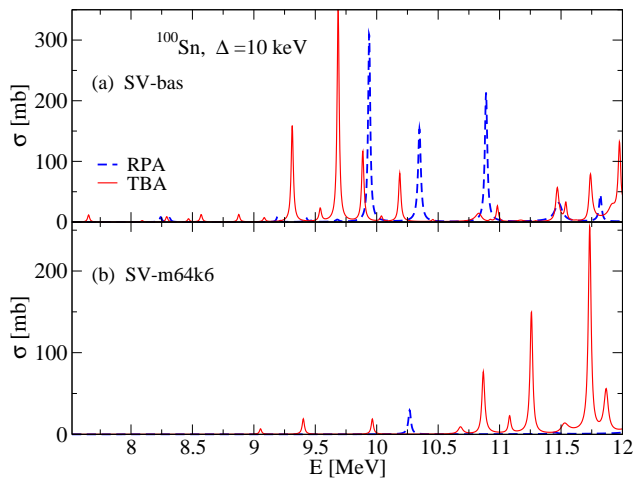}
\caption{\label{fig:100Sn_photo_delta10keV}
Same as in Fig.~\ref{fig:56Ni_photo_delta10keV} but for $^{100}$Sn.
There is only one RPA peak for the set SV-m64k6 below 12 MeV because
all the RPA strength for this set is above 12 MeV.}
}
\end{figure}
Figure \ref{fig:100Sn_photo_delta10keV} shows low-energy dipole
spectra for the proton rich $^{100}$Sn.  The situation, little
fragmentation and mainly down shift through TBA, looks very similar to
the proton rich $^{56}$Ni. The reason is also similar, namely the fact
that in this $N=Z$ situation the spectrum of low-lying $ph$ states is
very dilute. Neutrons are well bound thus having large level spacing
and separation energy. Protons have small separation energy, but the
Coulomb barrier produces still localized, quasi-discrete levels in the
continuum which also reduces the spectral density.  In case of
SV-m64k6 we have a curious effect from the plotting window.  We see in
TBA a group of peaks at the upper end of the displayed spectrum and
cannot spot the RPA ancestors. Note that these reside at somewhat
higher energies just outside the plotting window. What remains the
same as before is the typical difference that the SV-bas spectrum
reaches deeper down than that for SV-m64k6.

\begin{figure}\centering\footnotesize
\parbox[t]{8.5cm}{\centering
\includegraphics[width=8.5cm]{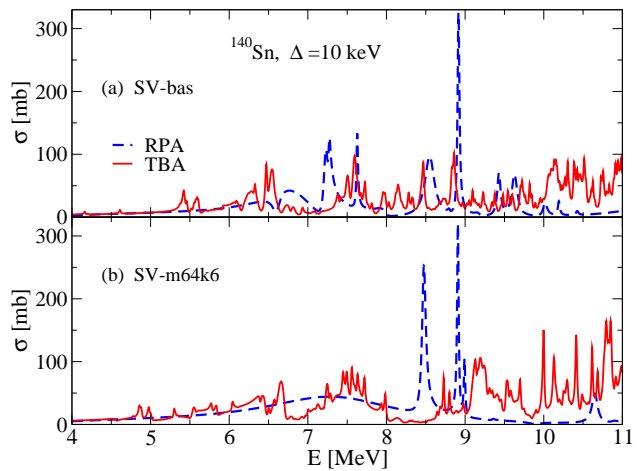}
\caption{\label{fig:140Sn_photo_delta10keV}
Same as in Fig.~\ref{fig:56Ni_photo_delta10keV} but for $^{140}$Sn.}
}
\end{figure}
As complement to $^{100}$Sn, Fig. \ref{fig:140Sn_photo_delta10keV}
shows low-energy dipole spectra for the very neutron rich Sn isotope
$^{140}$Sn. Much different than for $^{100}$Sn, TBA develops here a
strong bias toward spectral fragmentation leaving at the end basically
a structureless broad distribution over the whole energy range shown.
This is typical for nuclei toward the neutron drip line. Loosely bound
neutrons produce already at mean-field level a high density of states
which together with the phonons in TBA make up a considerably high
density of complex configurations delivering eventually these
basically flat distributions.  It is, furthermore, interesting to note
that, unlike previous examples, the spectra of SV-bas and SV-m64k6
cover the same energy range. This, again, happens already at RPA
level. It serves as a warning that some ``rules'' observed in stable
nuclei may fail in very exotic regions.


\section{Conclusions}

We investigated in the present publication the trends of
giant resonances as well as the behavior of the low-lying electric
dipole strength along the chains of Ni and Sn
nuclei, spanning from from proton rich to very neutron isotopes.
Only double magic Ni and Sn nuclei were considered.

The only exception is $^{140}$Ni which has only proton magic number.
But our HFB with unrestricted symmetry calculations show that even this nucleus
is spherical in the ground state and that the pairing correlations
have very little effect on the properties of its excited states.
As theoretical tool we use a fully self-consistent treatment within
    the random-phase approximation (RPA) extended by particle-phonon
    coupling to account for many-body correlations. Our calculations
  are based on a recently developed version of the time-blocking
  approximation
which allows an optimized selection of phonons and guaranties
a fast convergence.
This is important for a consistent
  treatment for all nuclei within the chains. The calculations were
  based on the Skyrme-Hartree-Fock (SHF) energy functional. We
  considered actually two previously adjusted SHF parametrizations
  with comparable quality in ground state properties, but different
  properties concerning giant resonances in order to explore the span
  of possible predictions. The two chains of doubly magic nuclei
  includes isotopes were experimental data are available and
  which serve nicely as benchmark for our approach. We then
    extended confidently our calculations to very exotic Ni and
  Sn isotopes which are relevant for astro-physical reaction
  chains.

  The effects for giant resonances are the same for all isotopes
  along the chains and as in previous explorations of stable
  nuclei. The phonon coupling mainly introduces an energy dependent
  broadening of the spectral distributions delivering realistic
  pattern close to experimental distributions.  There is also a small
  down-shift (0--1 MeV) of the average peak position.

  We also investigated in detail the low-lying electric dipole
  strength often summarized under the label pygmy dipole
    resonances (PDR). Here the phonon coupling can give rise to
  a remarkable shift of the low energy states, redistribution of
    strength, and often strong fragmentation of the already much
  reduced dipole strength. These effect are particularly
    pronounced for the very neutron rich nuclei amounting there to
    qualitative changes of the dipole spectra.  On the other side of
    proton rich states, phonon coupling has little influence because
    the density of low-energy phonon states is much lower.

As in many previous investigations, we do not observe any
collective behavior in the regime of low-energy isovector dipole states. We
  see rather a dense sequence of states with different internal
  structure.  The transition densities of these low-lying states show
for the proton rich $^{48}$Ni isotope proton dominated tails and for
the neutron rich isotopes neutron dominated tails \cite{Lyu19}.
This has, from our point of view, a
simple origin: In $^{48}$Ni the last occupied proton
shell includes $sp$ states with higher spins than the closed core. The
distribution of those states reaches more far outside giving rise to
the observed transition densities. The same is true for the neutron
rich isotopes were we observe tails which are neutron dominated.

The situation is different for the low-lying isoscalar electric dipole states.
Here the major part of the $1\hbar\omega$ $ph$-strength is removed
into the spurious state but due to the attractive isoscalar interaction
an appreciable amount of the $3\hbar\omega$ $ph$-strength is shifted
into the low-lying states. So if one is looking for some collectivity
in this regime the isoscalar electric dipole might be a candidate.
Further detailed analysis of the low-lying dipole spectra is presently underway.

\begin{acknowledgments}
N.L. and V.T. acknowledge financial support from the
Russian Science Foundation (Project No. 16-12-10155).
Research was carried out using computational resources
provided by Resource Center ``Computer Center of SPbU''.
This work had been supported also by the DFG
(contract Re322-13/1).
\end{acknowledgments}

\appendix

\section{Numerical considerations}
\label{app:numerics}

\subsection{Single-particle basis in TBA calculations.}
\label{app:sp-basis}

The size of $sp$ space is a compromise. It needs to be large
enough to produce converged results and we want to have it as low as
possible to render calculations affordable.  In our previous
calculations we checked the dependence of the results on the maximum
$sp$ energy. For light ($^{16}$O) and medium mass nuclei ($^{40,
  48}$Ca) we found saturation at $\ve^\mathrm{sp}_\mathrm{max}= 500$
MeV, and for heavy nuclei ($^{132}$Sn and $^{208}$Pb) at
$\ve^\mathrm{sp}_\mathrm{max} = 100$ MeV.  The present test cases of
Ni and Sn isotopes lie in between, they reach deep into regimes of
exotic nuclei, and they explore fine structure in the PDR region.
Thus the $sp$ basis has to be inspected again.  For computation of
spectra of giant resonances, we find again that an $sp$ basis limited
to $\ve^\mathrm{sp}_\mathrm{max} = 100$ MeV is sufficiently large.
The calculations of dipole spectra in the low-energy regions with
resolution $\Delta = 10$ keV requires to increase in the $sp$-basis.
We find that an upgrade from $\ve^\mathrm{sp}_\mathrm{max} = 100$ to
200 MeV suffices. A further increase in $\ve^\mathrm{sp}_\mathrm{max}$
to 500 MeV almost does not affect the results.
What $sp$ angular momentum is concerned,
saturation was reached in the TBA calculations when going up to
$l^{\mbsu{sp}}_{\mbsu{max}}$ =17 in all cases.

\begin{figure}[h]
\includegraphics[width=\linewidth]{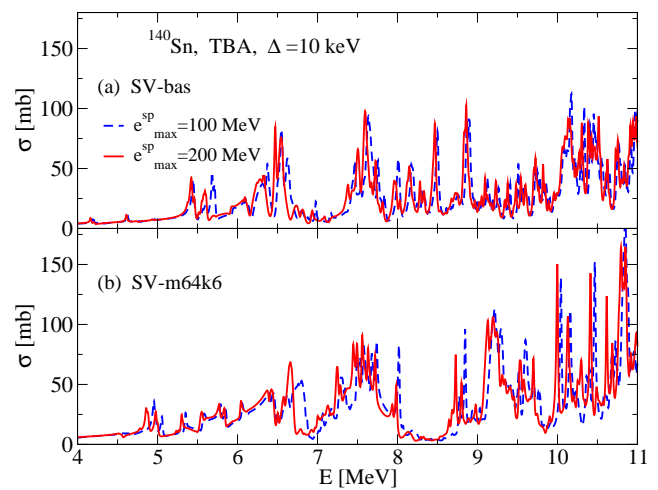}
\caption{\label{fig:140Sn_photo_espmax_effect_d10}
The effect of the $sp$-basis dimension to the cross section $\sigma(E)$
for PDR in $^{140}$Sn obtained in TBA with the EDF parameter sets SV-bas and
SV-m64k6 for $\ve^\mathrm{sp}_\mathrm{max} = 100$ and 200 MeV.
$\Delta=10$ keV.
}
\end{figure}
The effect of choice of $sp$
basis on the fine structure of the PDR strength is demonstrated in
Fig.~\ref{fig:140Sn_photo_espmax_effect_d10} for the case of
$^{140}$Sn. We still see tiny modifications when stepping from
$\ve^\mathrm{sp}_\mathrm{max} = 100$ MeV to 200 MeV. Nothing is seen
at plotting resolution for higher $\ve^\mathrm{sp}_\mathrm{max}$.  The
choice $\ve^\mathrm{sp}_\mathrm{max} = 200$ MeV for low-energy spectra
is clearly at the safe side. Robust moods would even have accepted the
lower choice of 100 MeV.

\subsection{Effect of box size}
\label{app:box}

The box size was $R_{\mathrm{box}} = 18$ fm for all the nuclei.
Such a value gives reliable results for calculations with $\Delta=400$ keV.
To investigate the influence of the box size on the fine structure
of the E1 strength we have calculated this strength in RPA and TBA
with $\Delta=10$ keV for $R_{\mathrm{box}} = 18$ and 30 fm.
These calculations were made for $^{100}$Sn having small $S(p)$,
$^{140}$Sn having small $S(n)$, and $^{208}$Pb having large
both the separation energies.
For each of these nuclei the RPA results for $R_{\mathrm{box}} = 18$ and
30 fm coincide therefore corresponding figures are of no interest.
\begin{figure}
\centering\footnotesize
\parbox[t]{8.5cm}{\centering
\includegraphics[width=8.5cm]{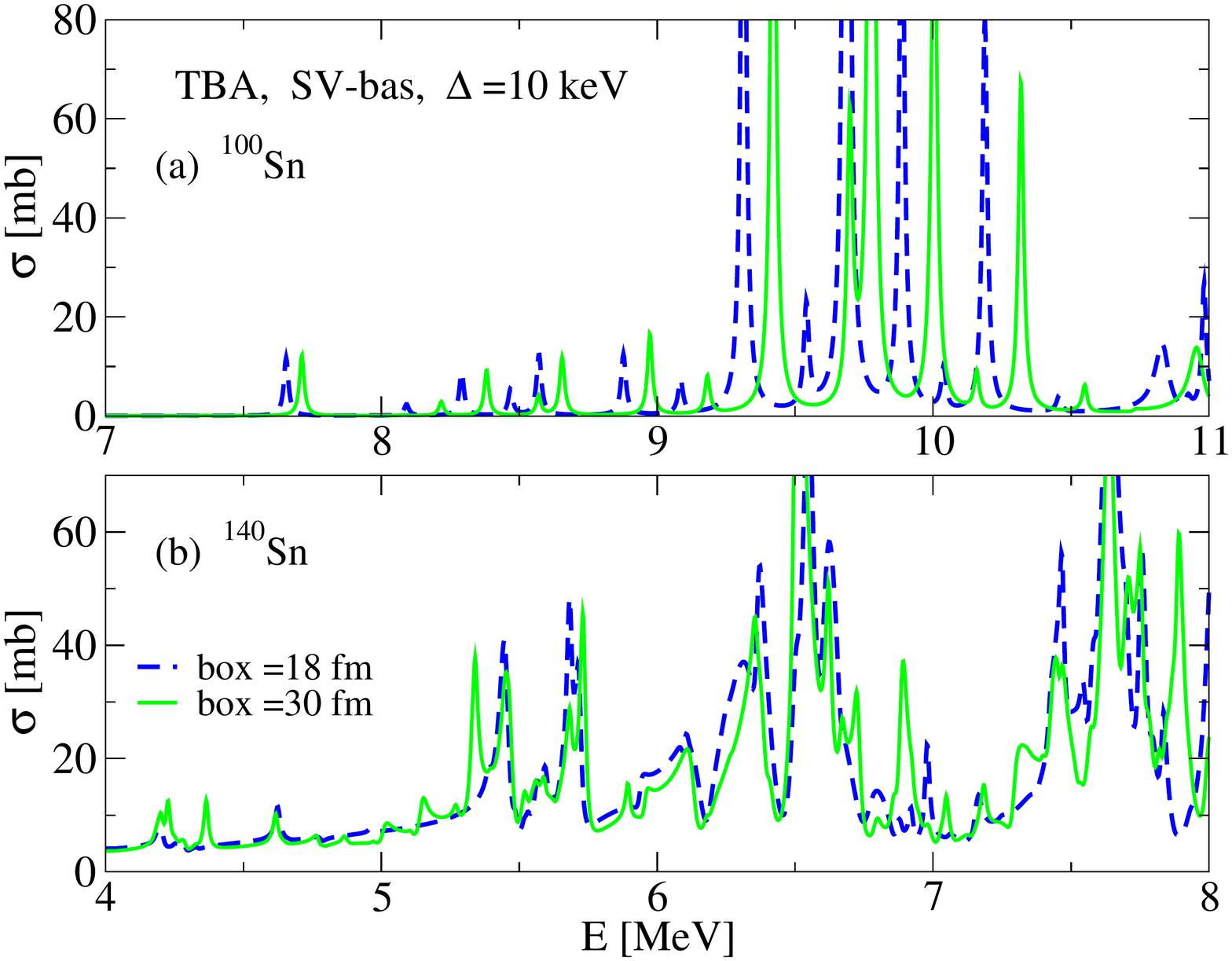}
\caption{\label{fig:100_140Sn_photo_bas_box_effect_d10}
The box-size dependence of the photo-absorption cross section
in $^{100,140}$Sn in the low-energy region calculated in the TBA with
$\Delta=10$ keV for the parameter set SV-bas.}}\hfil
\parbox[t]{8.5cm}{\centering
\includegraphics[width=8.5cm]{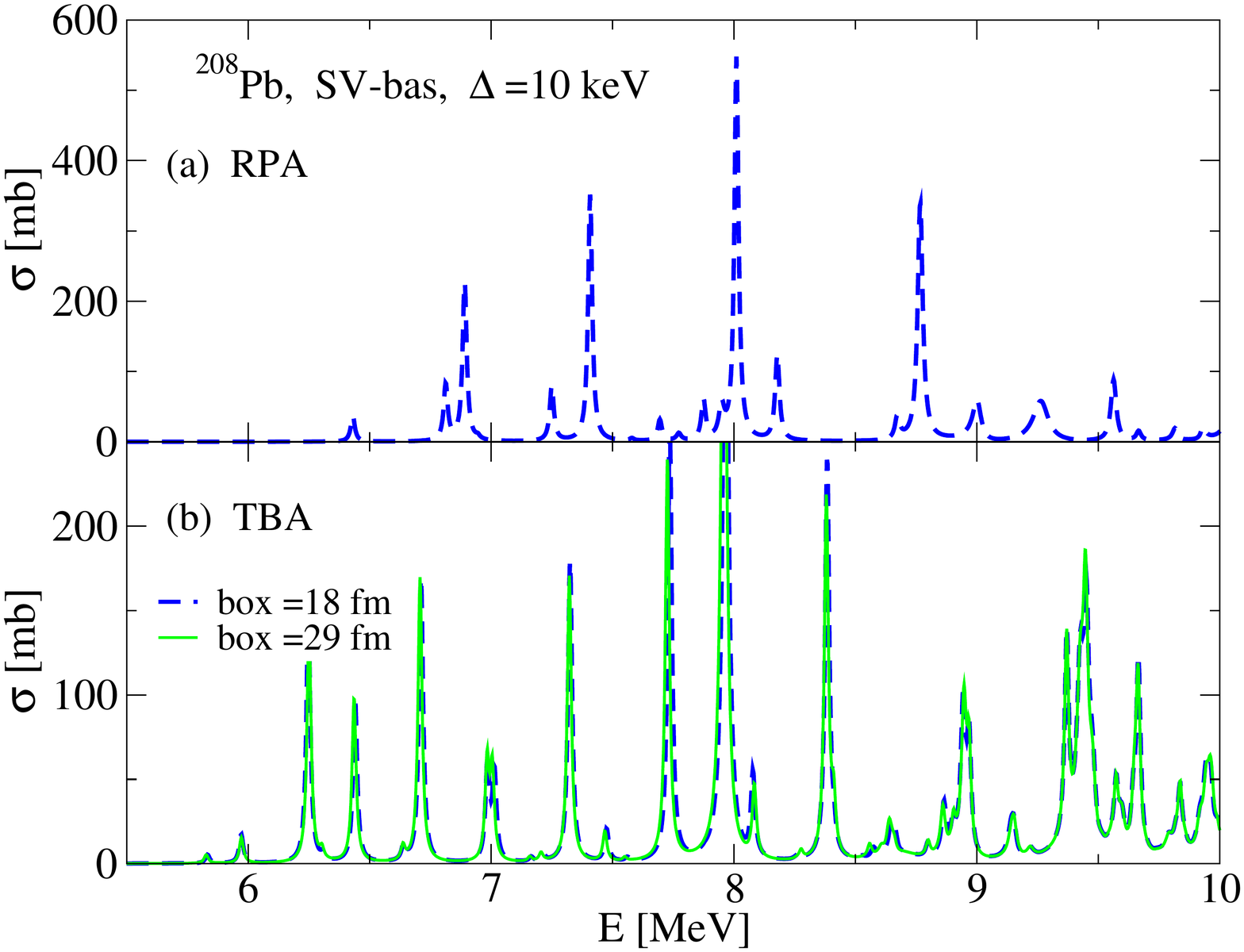}
\caption{\label{fig:208Pb_photo_bas_box_effect_d10}
Same as in Fig.~\ref{fig:100_140Sn_photo_bas_box_effect_d10}
but for $^{208}$Pb.}}
\end{figure}
The box-size dependence of the TBA results is shown in
Figs.~\ref{fig:100_140Sn_photo_bas_box_effect_d10} and
\ref{fig:208Pb_photo_bas_box_effect_d10}.  For $^{100}$Sn, where
$S(p)$ is small, almost twofold increase in the box size does not
change the form of the TBA E1 strength in the pygmy region and only
slightly, by 0.1 MeV, shifts all the peaks to high energies: see
Fig.~\ref{fig:100_140Sn_photo_bas_box_effect_d10}(a).  This shift of
the peaks is due by a similarly small shift in the phonon energies
when changing box size.  Since neither $\ve^\mathrm{sp}_\mathrm{max}$,
no $R_{\mathrm{box}}$ variation change the number of peaks, we
conclude that for this case of $^{100}$Sn all the peaks are of a
physical nature.  The same is found for the fine structure of the PDR
in $^{208}$Pb obtained in TBA calculations with $\Delta = 10$ keV. It
does not depend on the box size, at least for $R_{\mathrm{box}} \geq
18$ fm. We expect that this holds also for all other stable nuclei.

A different picture is seen for $^{140}$Sn having very small neutron
separation energy $S(n)$: see
Fig.~\ref{fig:100_140Sn_photo_bas_box_effect_d10}(b).  There is a
noticeable (not very large) box dependence of the TBA results for the
PDR fine structure.  The effect is better seen in the low-energy
region where the strength structure is not very complex.  The main
effect is an appearance of additional small peaks.  The nature of
these peaks may be explained if we take into account that, above the
nucleon threshold, the nucleus in a box may be approximately
considered as a Fermi gas in the box with infinite walls.  The
strength from discrete RPA (as used for the definition of phonons) for
such a system has peaks whose number on a given energy interval is
proportional to the size of the box.  One can see a similar periodic
structure in Fig.~\ref{fig:100_140Sn_photo_bas_box_effect_d10}(b) for
energies below 5.4 MeV.  Here small additional TBA peaks arise because
of the discrete nature of phonons in the complex configurations.  The
TBA peaks at $E =$ 5.4 and 5.7 MeV are probably real physical peaks.
This explanation is, of course, very simplified.  The main conclusion
that can be drawn from this consideration is that one can not use a
very small $\Delta$ in TBA calculations for a nucleus with a low
nucleon separation threshold.  Probably, a minimal value for the
exotic Sn isotopes is $\Delta = 20$ keV.

\bibliographystyle{apsrev4-1}
\bibliography{TTT}

\end{document}